\documentclass[twocolumn,showpacs,preprintnumbers,amsmath,amssymb]{revtex4}

\usepackage{graphicx}
\usepackage{dcolumn}
\usepackage{bm}
\usepackage{amssymb}
\usepackage{amsmath}
\usepackage{latexsym}

\begin{document}

\title{Medium effects on the selection of sequences folding \\
into stable proteins in a simple model}%

\author{You-Quan Li}\email{yqli@zimp.zju.edu.cn}
\author{Yong-Yun Ji}
\author{Jun-Wen Mao}
\author{Xiao-Wei Tang}
\address{Department of Physics, Zhejiang University, Hangzhou 310027, P.R. China.} %

\date{\today}%

\begin{abstract}
We study the medium effects
on the selection of
sequences in protein
folding by taking account
of the surface potential in
$HP$-model. Our analysis on
the proportion of H and P
monomers in the sequences
gives a direct
interpretation that the
lowly designable structures
possess small average gap.
The numerical calculation
by means of our model
exhibits that the surface
potential enhances the
average gap of highly
designable structures. It
also shows that a most
stable structure may be no
longer the most stable one
if the medium parameters
changed.

\end{abstract}


\maketitle

Proteins are known to play a virtual role in the structure and
functioning of all forms of life, and the protein folding problem
is one of the most fundamental and still unsolved problems.
Composed of a specific sequence of amino acids, each protein is
folded into native structure (a particular 3-dimensional shape)
that determines its biological function and it is widely believed
that for most single domain proteins, the native structure is the
global free-energy minimum\cite{1}. The amino-acid sequence alone
encodes sufficient\cite{1} information to determine its 3-d
structure. Theoretical studies on protein sequence and structure
include molecular dynamical simulation\cite{2} and lattice
model\cite{3}. The latter has absorbed much attention\cite{4,5}
while the former takes much CPU even on huge computers\cite{2}.

For the naturally occurring
varieties of amino acids
can be classified\cite{6}
as either of hydrophobic(H)
or of polar(P), a
HP-lattice model to
interpret protein folding
was introduced\cite{4}.
Based on the called
standard HP model, 27
monomers occupying all
sites of a cubic\cite{5},
Li et al.\cite{7}
introduced the
designability to show that
potentially good sequences
are those with a unique
ground state separated by a
large gap from the first
excited state. By defining
the designability of a
structure as the number of
sequences that possess the
structure as their unique
lowest-energy state, they
found that the structures
differ drastically in their
designabilities. The
sequences that design the
highly designable
structures are
thermodynamically more
stable\cite{7,8}. Studies
on the designability for a
larger lattice
model\cite{9} and for an
off-lattice model\cite{10}
showed the similar results.
For many-letter models, the
different parameters gave
different results: Buchler
et al.\cite{11} got that
the designability of the
structure depends
sensitively on the size of
the alphabet, and Li et
al.\cite{12} achieved that
the designability of the
structure is not sensitive
to the alphabet size when a
realistic interaction
potential(MJ matrix) is
employed. Ejtehadi et al.
found that if the strength
of the non-additive part of
the interaction potential
becomes larger than a
critical value, the degree
of designability of
structures will depend on
the parameters of the
potential\cite{13}.

Since useful features
concerning to the protein
folding and their stability
can be explored on the
basis of lattice model, it
will be worthwhile to study
the effect of media on
protein folding properties.
In this letter, we consider
the medium effects by
introducing different
parameters that
characterize various
concentrations of medium
solution. Our results give
some answers to the
following questions.
Namely, are those sequences
associated with highly
designable structures
universally good? how do
they vary depending on
media\cite{14} where the
protein is placed?


We investigate the effects
of media upon the category
of highly designable
protein sequences, which
will undoubtedly provide a
clue to understand the
variations in the nature
selection of protein
species caused by media
where the protein lives.
For this purpose, we must
reconstruct the original HP
model by introducing
potential parameters to the
monomers at protein's
surface. The protein is
figured as a chain of beads
occupying the sites of a
lattice in a self-avoiding
way, so our model
evaluating the energy of a
sequence folded into a
particular structure reads,
\begin{equation*}
H=\sum_{i<j}E_{\sigma_i\sigma_j}\delta_{|r_i-r_j|,1}(1-\delta_{|r_i-r_j|,1})
+\sum_{{r_j}{\in}S}U_{r_j}\delta_{\sigma_j,P}
\end{equation*}
where i, j denote for the
successive labels of
monomers in a sequence,
$r_i$ for the position (of
the $i$-th monomer) on the
lattice sites, and
$\sigma_i$ refers H or P
corresponding to
hydrophobic or polar
monomer. Here the Kronecker
delta notation is adopted,
i.e., $\delta_{a,b}=1$ if
a=b but $\delta_{a,b}=0$ if
$a\ne b$. As the
hydrophobic force\cite{6}
drives protein to fold into
a compact shape with more
hydrophobic monomers inside
as possible, the $HH$
contacts are more favorite
in this model, which can be
characterized by choosing
$E_{PP}=0$, $E_{HP}=-1$,
and $E_{HH}=-2.3$ as
adopted in Ref.\cite{7}. In
order to include the
effects caused by the
protein's surrounding
medium that is relevant to
salt concentration\cite{14}
of a solution where the
protein is placed, we
introduce $U_V$, $U_E$, and
$U_F$ to represent the
attractive potentials in
the protein surface for
polar (hydrophilic)
monomers at vertices,
edges, or face centers
respectively. These
attractive forces arise
from the medium (solution)
to the hydrophilic
monomers. Since we are not
able to deal with a sphere
surface in present lattice
model, we consider
different weights at the
surface, saying
$U_{\tau}=-\gamma_{\tau}V$.
If
$\gamma_V=\gamma_E=\gamma_F\ne
0$, no any new results
occur in comparison to the
result that Li et al. had
studied. This is because
the core in the cubic of
the 27-site model always
contains a hydrophobic
core, which implies that
the surface potentials
merely cause a global shift
in energy spectrum of the
27-site model if we impose
an equal weights on a
vertex, edge as well as
center of a face. We then
investigate several cases
of non-vanishing,
$\gamma$'s later on.

\begin{figure}
\includegraphics[width=0.32\textwidth]{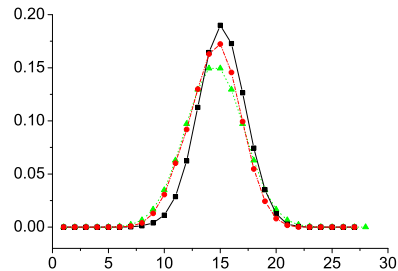}%
\caption{\label{fig:binary} Comparison of distributions for binary
arrangement (green dot line), the lowly designable structures (red
dash-dot line), and the highly designable structures (black solid
line) respectively.}
\end{figure}

It has been noticed\cite{7}
that some structures can be
designed by a large number
of sequences, while the
others can be designed by
only few sequences. The
designability of a
structure is measured by
the number($N_s$) of
sequences that take the
given structure as their
unique ground state, as was
first introduced by Li et
al.\cite{7}. Additionally,
structures differ
drastically according to
their designability, i.e.,
highly designable
structures emerge with a
number of associated
sequences much larger than
the average ones. For a
particular sequence, the
energy gap $\delta_s$ is
the minimum energy needed
to change its ground-state
structure into a different
compact structure. The
average energy gap
$\bar{\delta}_s$ for a
given structure is
evaluated by averaging the
gaps over all the $N_s$
sequences that design that
structure. The structures
with large $N_s$ have much
larger average gap than
those with small $N_s$, and
there is an apparent jump
around $N_s=1400$ in the
average energy gap. This
feature was first noticed
by Li et al.\cite{7} in the
medium-independent  HP
model, thus these highly
designable structures are
thermodynamically more
stable and possess
protein-like secondary
structures into which the
protein sequences fold
faster than the  other
structures\cite{7}. To
interpret this feature, we
calculate the  average
distribution of the number
of hydrophobic monomers for
the highly designable
structures and for the
lowly designable structures
respectively. We plot these
two distributions together
with the pure mathematical
binary arrangement
distribution in Fig.~\ref{fig:binary} where
all distributions are
normalized to unit.
Clearly, the distributions
for highly designable
structures shift toward the
larger number of
hydrophobic monomers in
comparison to the
mathematical distribution.
This leads to a lower
energy scale   because the
more hydrophobic monomers
there are, the lower their
energy will be. Oppositely,
the distribution for lowly
designable structures shift
toward the small number of
hydrophobic monomers in
comparison to the
mathematical distribution,
which causes a higher
energy. This may interpret
that the lowly designable
structures possess small
average gap.

Although the choices of
$E_{PP}=0$, $E_{HP}=-1$,
and $E_{HH}=-2.3$ adopted
in Ref.\cite{7} fulfil the
principle that the major
driving force for protein
folding is the hydrophobic
force, the difference
between the H-H contacts
occurring inside protein
and that occurring at surface was disregarded. Therefore, to explore the designability
affected by the medium surrounding the protein, the application of surface
potential in our model becomes inevitable. We pointed out in the above that
the 26 monomers are on the surface for 27-site model, which gave trivial
result for uniform weights to the surface potential. On the other hand,
increasing the number of the lattice sites will make the model beyond the
calculation capacity of nowadays computers. However, after some further
tuning the original model, we are able to obtain nontrivial and interesting
results. First, we consider a ``cubic shape approximation" by imposing different
potential weights:
${\gamma}_V=7/8$, ${\gamma}_E=6/8$,
and ${\gamma}_F=4/8$, which come from the different interfaces
between the medium solution and the monomers at vertex, edge and the face centre
respectively. For this parameter choice, we find there are 17 more sequences
possessing unique ground state regardless of the magnitudes of $V$
(ranging from  0.1 to 2.1) though they do not possess unique ground states in the model studied
by Li et al where the effect of medium was neglected\cite{7}. Our calculation further
exposes that 14 of those 17 sequences mainly belong to the highly designable
structures, and have relatively larger energy gap. We analyse all the 17 sequences,
and find that the 14 ones can be related to each other by a single mutation, which
implies that they belong to the ``neutral island" suggested by Trinquier et al.\cite{15}.
These results confirm that protein structures are selected in nature because they
are readily designed and stable against mutations, and that such a selection
simultaneously leads to thermodynamic stability and foldability. Thus, a key
point to understand the protein-folding problem is to understand the emergence
and the properties of highly designable structures.

\begin{figure*}
\includegraphics[width=0.28\textwidth]{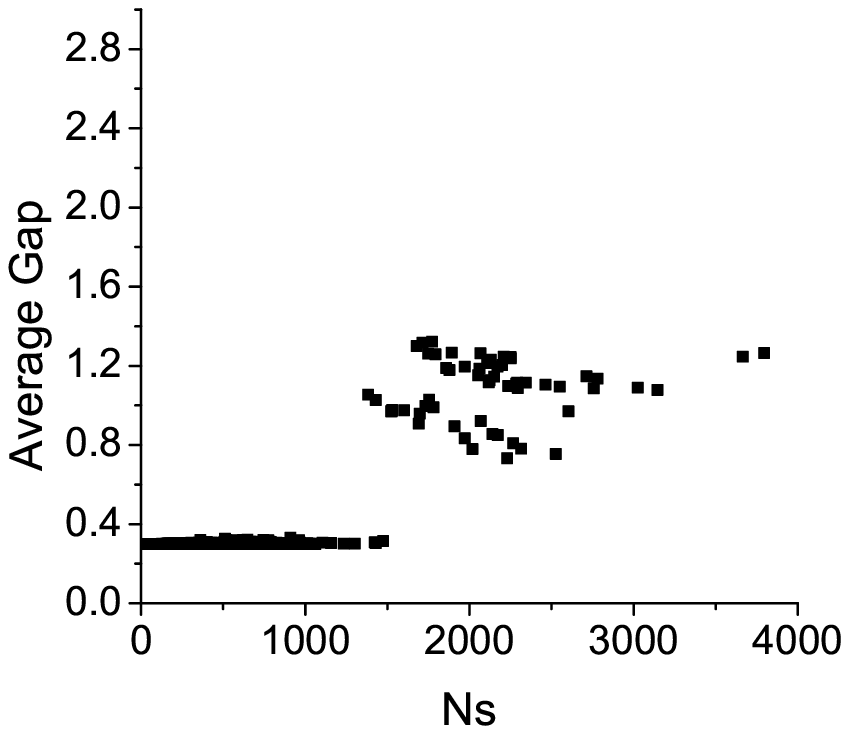}
\includegraphics[width=0.28\textwidth]{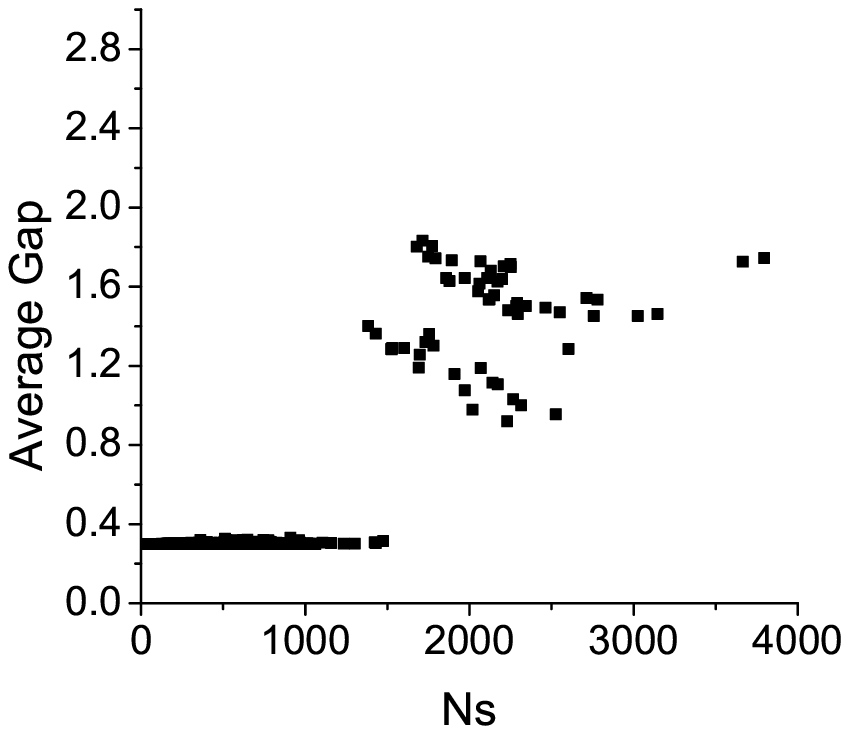}
\includegraphics[width=0.28\textwidth]{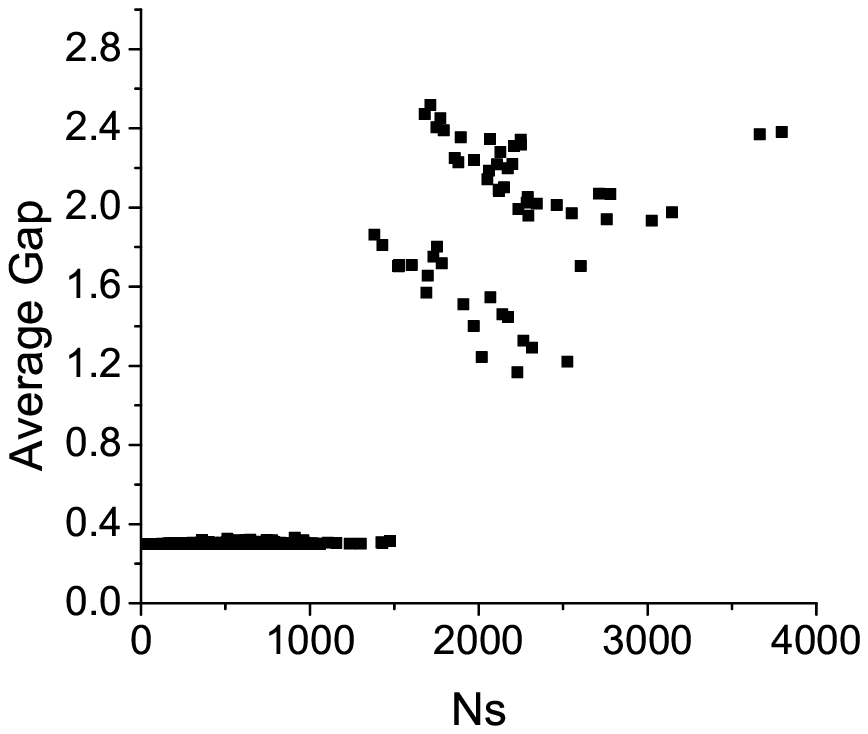}
\caption{\label{fig:average}Average
gap of structures versus $N_s$ of the structures in the case of
${\gamma}_V=7/8$, $\gamma_E=6/8$, $\gamma_F=0$ for (a) $V=0.0001$,
(b)$V=0.9$, and (c)$V=2.1$, respectively.}
\end{figure*}

\begin{figure*}
\includegraphics[width=0.28\textwidth]{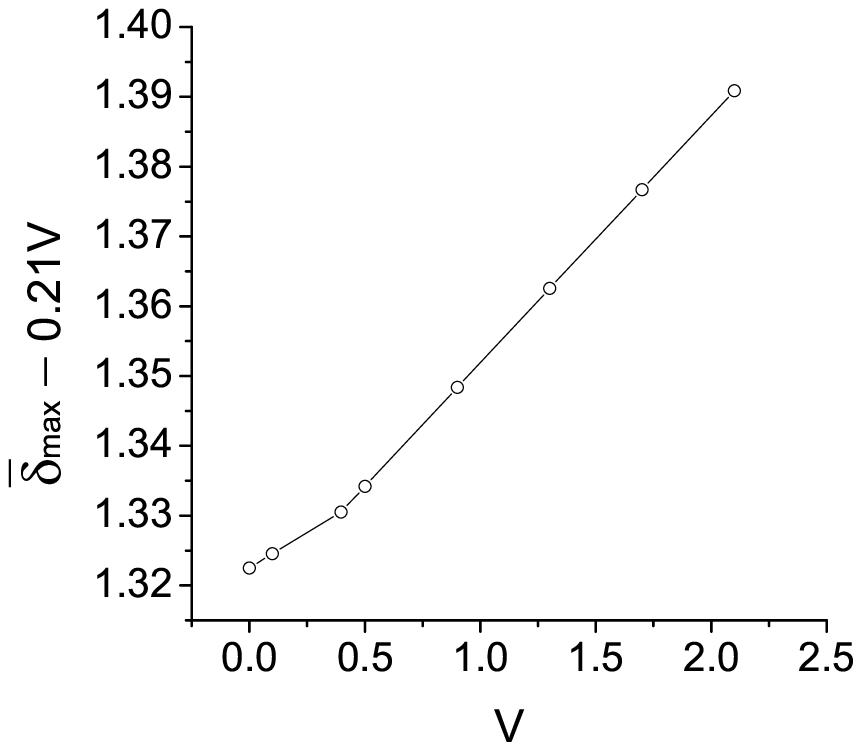}%
\includegraphics[width=0.28\textwidth]{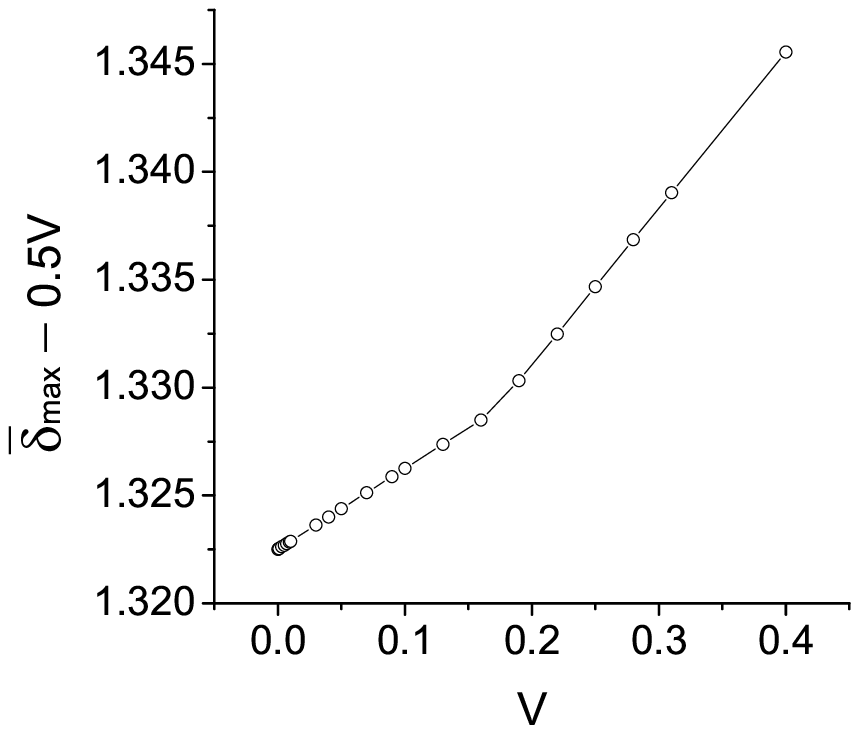}
\includegraphics[width=0.28\textwidth]{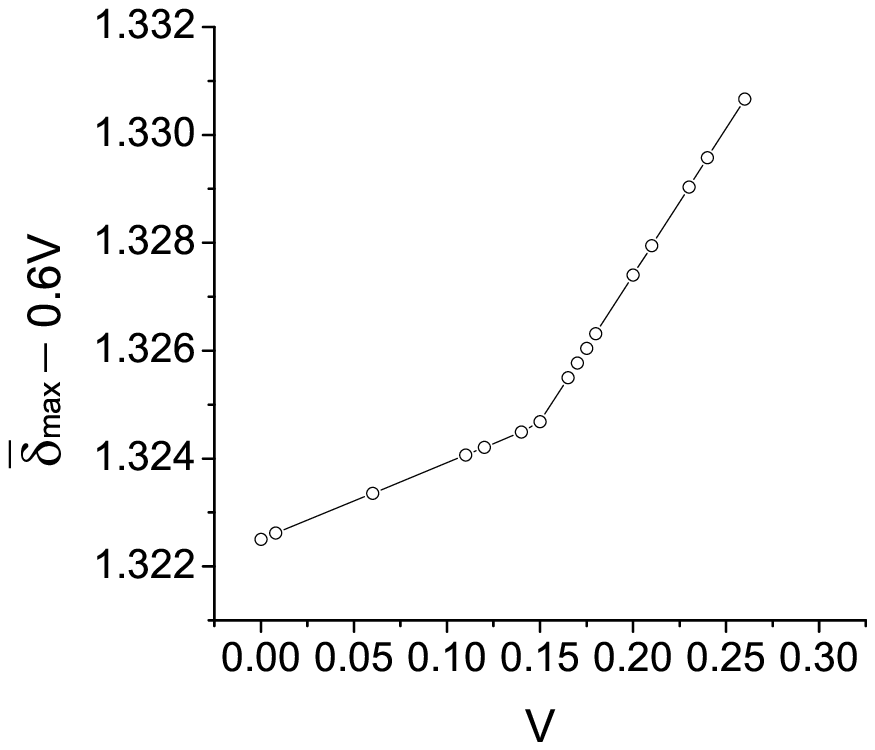}
\caption{\label{fig:largest}
The largest average gap
$\bar{\delta}_{max}$ versus
the parameter $V$: (a) for
$\gamma_V=7/8$,
$\gamma_E=6/8$,
$\gamma_F=4/8$; (b)for
$\gamma_V=7/8$,
$\gamma_E=6/8$,
$\gamma_F=0$;
(c) for $\gamma_V=1$, $\gamma_E=1$, $\gamma_F=0$ case.}
\end{figure*}

\begin{figure*}
\includegraphics[width=0.28\textwidth]{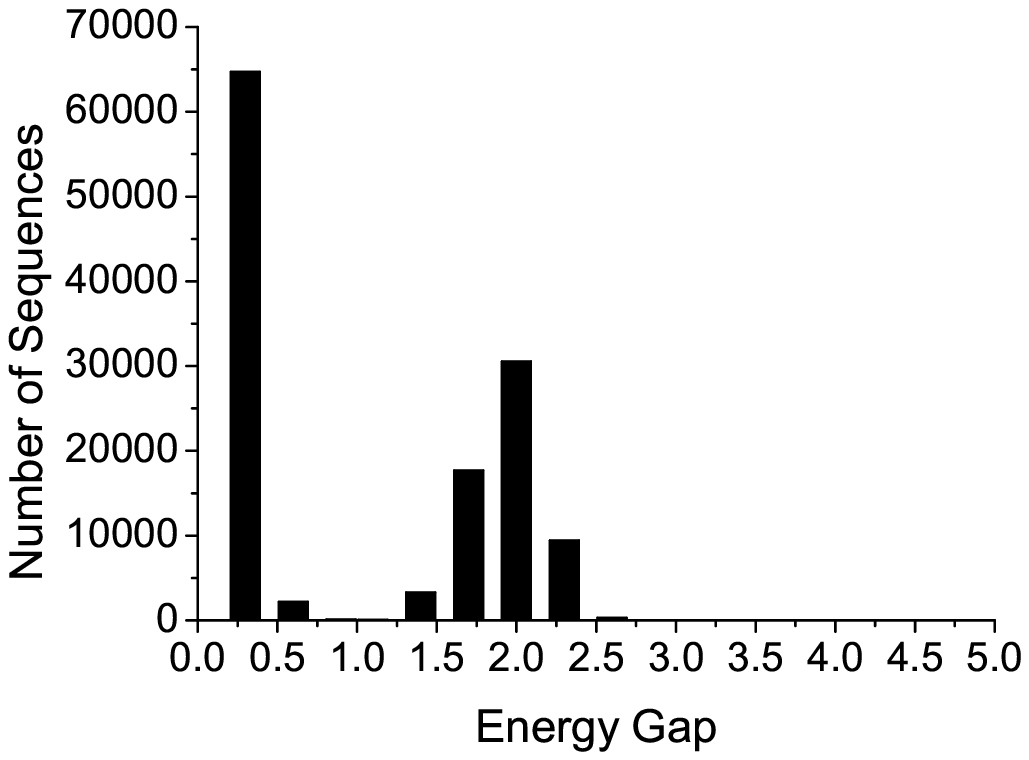}
\includegraphics[width=0.28\textwidth]{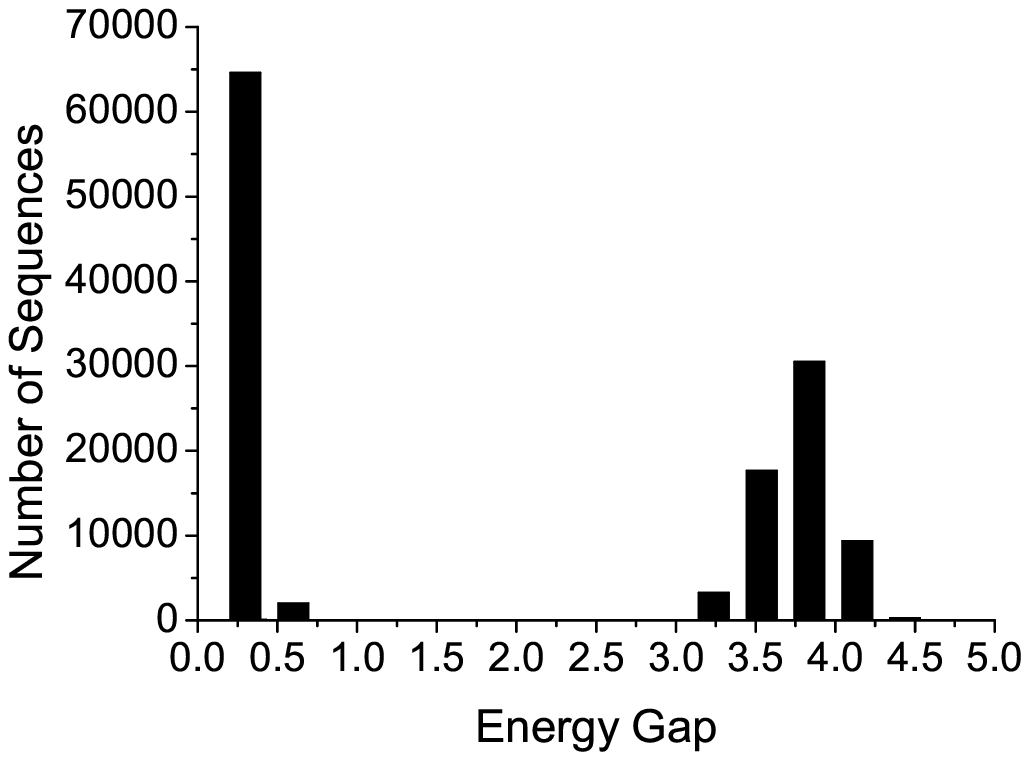}
\caption{\label{fig:histogram}The
histogram for the number of
sequences versus the energy
gap for the 60 high
designable structures in
the absence of medium
(left); and in the presence
of medium $\gamma_V=7/8$,
 $\gamma_E=6/8$,$\gamma_F=0$, $V=2.1$(right).}
\end{figure*}

The second parametrization is to consider
$\gamma_V=7/8$,
$\gamma_E=6/8$, and $\gamma_F=0$, which models
a protein with 7 monomers at the inside while 20 ones
at surface. In this case, we find  there are 48 more
sequences possessing unique ground state for a wider
range of magnitudes  of $V$ (from 0.0001 to 2.1),
which, however, have none unique ground states in the
case of Li et al.\cite{7}. Whereas, only one sequence
designs the highly designable structure while
the other 47 sequences design lowly designable
structures. All the energy gaps of those new sequences
are found to be $V/8$.
Since the ratio of the
numbers of the monomers at
surface to that at the
inside is of order 1 in
natural proteins\cite{8},
and the ratio in our model
is 26:1 in first case but
is 20:7 in the second case,
the latter case ought to be
closer to the usual natural
proteins. Fig.~\ref{fig:average} shows the
average energy gap for
different potential
parameters. Clearly, the
surface potential enhances
the average gap of highly
designable structures,
which illustrates that the
highly designable
structures selected by
nature are more stable in
proper media than in
``vacuum". Recent
experiment\cite{16}
revealed that the
additional stability of a
thermophilic protein comes
from just a few residues at
the protein surface. Thus
our theoretical results may
evoke more attention to the
dependence of stability on
medium effects in further
model studies.

We calculate the case by assuming the potentials at
the vertices and at edges with the same weights,
i.e., $\gamma_V=1$,  $\gamma_E=1$, and $\gamma_F=0$.
We find that there is no sequence beyond those of Ref.\cite{7} to take
the highly designable structures. Just like the result in Ref.\cite{14}, there are
also 60 structures that possess large average gap. When we take
account of the effects of medium, the average gap for highly
designable structures increase apparently as the potential
parameter increases, but the average gap of lowly designable
structures does not change much. In all the aforementioned
cases, the average gap of a single highly designable structure
increases linearly with
respect to the increase of
$V$. Furthermore, we find
the structure with largest
average gap is not fixed
for all potential
parameters. Crossings
between energy levels
always take place when the
potential parameter
changes. It is therefore
worthwhile to point out
that the gains of stability
for distinct structures
vary, and the most stable
protein structure in one
surrounding medium maybe no
more the most stable one in
another medium. The plots
of the largest energy gap
versus the parameter $V$
are shown in Fig.~\ref{fig:largest}
respectively for the three
cases of the weights
$\gamma$'s discussed in the
above. In order to show an
apparent change for eye's
view, we have set the value
of the vertical axis in
Fig.~\ref{fig:largest} to be the largest
average gap minus $0.21V$,
$0.5V$, and $0.6V$ for the
cases (a), (b), and (c),
respectively. In each case
is there a critical value
of $V$ across which the
plot transits from a strait
line to another strait
line. The critical values
of $V$ differ in different
cases, but the largest
average gaps at the
transition point take the
same value
$\bar{\delta}_s=1.4137$.

We analyze all the
sequences  that design the
60 highly designable
structures respectively. In
the absence of medium,
${\gamma}_V={\gamma}_E={\gamma}_F=0$,
the energy gaps
${\delta}_s$ of those
sequences range from 0.3 to
2.6 (see Fig.~\ref{fig:histogram}). Almost
half of them have small
energy gaps (around 0.3).
In the presence of medium,
the energy gaps for most of
the sequences with larger
(over 1) energy gap rises
as parameter increases
while that for the
sequences with small energy
gap does not rises
apparently. For the cases
(a) ${\gamma}_V=7/8$,
${\gamma}_E=6/8$,
${\gamma}_F=4/8$, (b)
${\gamma}_V=7/8$,
 ${\gamma}_E=6/8$, ${\gamma}_F=0$, and
 (c) ${\gamma}_V={\gamma}_E=1$, ${\gamma}_F=0$,
 the increments in energy gaps are mainly $3V/8$, $7V/8$, and
 $V$
 respectively. Whereas, there are also a small portion of the
sequences whose energy gaps decrease in the medium, e.g., 276
sequences in the case
 ${\gamma}_V=7/8$,
 ${\gamma}_E=6/8$,
${\gamma}_F=4/8$.
Considering some particular
structures among the 60
highly designable ones, we
analyze the sequences that
design them. The energy gap
of the sequences with
larger energy gap will
mostly increase when the
sequence is placed in
medium, which leads to the
linear increment of average
gap. Our results also
illustrate that the
distribution shapes emerge
similar for those three
structures.  In addition,
the total number of
sequence in (b) is less
than in (c), but there are
much more sequences
possessing large energy gap
in (b) than in (c).

In summary, our simple analysis of the average
distribution of the number of hydrophobic monomers can
interpret that the lowly designable structures
possess small average gap. Our model study exhibits
that the surface potential enhances the average gap of
highly designable structures, which implies
that the highly designable structures selected by
nature are more stable in proper media than in
``vacuum". We obtained that the energy gap of the
sequences with larger energy gap will mostly
increase when the sequence is placed in medium, which
leads to the linear increment of average gap.
We also noticed that there is a critical value for the
parameter of the surface potential, which means that
a most stable structure may be no longer the most
stable one if the medium parameters changed. Since a
lot of studies have shown that several properties of
natural proteins can be captured by simple models,
our discussion in above may motivate people to model
the effect of medium on all theoretical studies  where
the medium potential was ignored.

This work is supported by NSFC No.10225419 and 90103022.


\end{document}